# Diminished spin-flip reflectivity in stacked multilayers with varying period thicknesses of Fe/Si by incorporating $^{11}B_4C$

A. Zubayer[1], N. Ghafoor[1], A. Devishvili[2], A. Vorobiev[2,6], A. Glavic[3], J. Stahn[3], T. Hanashima[4], J. Sugiyama[4], V.A. de Oliveira Lima[5], J. Birch[1], F. Eriksson[1]

1. Department of Physics, IFM, Linköping University, SE-581 83 Linköping, Sweden
2. Institut Max von Laue - Paul Langevin, 71 avenue des Martyrs, 38042 Grenoble; France
3. PSI Center for Neutron and Muon Sciences, 5232 Villigen PSI, Switzerland
4. Neutron Science and Technology Center, Comprehensive Research Organization for Science and Society (CROSS), Tokai, Ibaraki 319-1106, Japan
5. Jülich Centre for Neutron Science (JCNS-2), Forschungszentrum Jülich GmbH, 52425 Jülich, Germany
6. Department of Physics and Astronomy, Uppsala University, SE-75120 Uppsala, Sweden

Corresponding Author: Anton Zubayer, anton.zubayer@liu.se, +46760583232

**Abstract**

This study investigates the effects of $^{11}B_4C$ co-sputtering on the structural and optical properties of Fe/Si stacked multilayers, with a focus on neutron supermirror applications. X-ray and neutron reflectivity techniques reveal that $^{11}B_4C$ incorporation improves interface sharpness, reduces roughness, and enhances reflectivity for all multilayer periods. Neutron reflectivity measurements show reduced spin-flip intensities, while wafer-curvature measurements indicate a 50% reduction in internal stress, allowing for higher mechanical stability of the multilayers. These improvements are attributed to the amorphization of Fe layers, which also suppress the formation of structural and magnetic domains responsible for stress and spin-flip scattering. In contrast, the pure Fe/Si sample exhibits a persistent half-order Bragg peak, indicating residual antiferromagnetic coupling. The results demonstrate that $^{11}B_4C$ enhances neutron optics by reducing spin-flip effects, increasing reflectivity and polarization, and alleviating stress, enabling the use of polarizers at reduced external fields compared to pure Fe/Si multilayers. These findings establish $^{11}B_4C$ as a transformative material for advancing neutron supermirror technology, paving the way for more efficient, stable, and high-performance polarizers in next-generation neutron optics.

## Introduction

In the field of material science, neutron scattering plays a crucial role.[1,2] This technique relies on the use of neutron optics, guide, shape, polarize, and analyze neutron beams as they travel from their source to the detector.[3] These optics must be efficient in handling neutrons due to the inherently limited signal strength in neutron scattering experiments.[4] Minimizing neutron loss is therefore essential. Particularly, the ability of efficiently polarizing neutrons before they interact with a sample of interest, and then analyzing the polarized state after the interaction, is vital for reliable results, especially in fields such as magnetism research.[5] However, there are limitations in the current state-of-the-art Fe/Si multilayer material system for polarizing neutron optics. These limitations include reduced reflectivity, which can be attributed to the interface width between layers.[6–11] Additionally, the presence of magnetic impurities in the magnetic layers leads to spin-flip events,[12] which alter the neutron's spin state and consequently compromise the polarization purity of the neutron beam. In neutron reflectometry (NR) and polarized neutron reflectometry (PNR) spin-flip scattering arises from the perpendicular component of magnetization of the magnetic structure, often due to magnetic domains or coupling effects within the sample. Biquadratic or antiferromagnetic coupling between layers can introduce spin canting or antiferromagnetically aligned magnetic moments across the layers, resulting in a non-uniform magnetic configuration that scatters neutrons at half the scattering vector Q, thereby reducing polarization performance.[13,14]

To address these challenges, earlier studies incorporating $^{11}B_4C$ in periodic Fe/Si multilayers with only 10 or 20 number of periods have shown great improvements in two-state polarizing neutron reflectivity, magnetic softness, polarization, and the capability to reflect at higher scattering angles.[15] To build on our findings, here, a comprehensive study of the reflectivity in stacked multilayers with hundreds of periods and varying period thicknesses (150, 75, 37.5, and 25 Å) to mimic a supermirror with a large $m$ is conducted. We also examined the spin-flip behavior of Fe/Si + $^{11}B_4C$ multilayers, as no previous studies have assessed whether polarized neutrons retain their spin state or undergo flipping upon interaction.

A key challenge in achieving high $m$-value polarizers is the increased number of periods, which increases internal stress in the multilayer/supermirror. This stress can cause films to crack or delaminate during coating or after exposure to ambient conditions. Stress in stacked Fe/Si multilayers has been investigated, with the incorporation of $^{11}B_4C$ hypothesized to mitigate this issue by promoting amorphization of the Fe layers. Amorphous materials typically exhibit lower internal stress due to the absence of atomic order and reduced strain from lattice mismatches.[16] This stress reduction could enable the fabrication of thicker films and higher $m$-values, which are necessary for high-performance polarizing supermirrors.

In this study, Fe/Si and Fe/Si + $^{11}B_4C$ multilayers are compared in terms of reflectivity, spin-flip behaviour and stress. Two stacked multilayer samples were fabricated, each consisting of four sets of multilayers (A, B, C, and D), each set giving rise to a distinct primary Bragg peak, as shown in Figure 1. The only difference between the pure Fe/Si multilayer and the Fe/Si + $^{11}B_4C$ multilayer was the substitution of 15 vol.% of Fe and Si with 15 vol.% of $^{11}B_4C$. Additionally, two multilayers, one Fe/Si and one Fe/Si + $^{11}B_4C$ , with a multilayer period $\Lambda$ = 25 Å, a layer thickness ratio of $\Gamma$ = 0.5 and N = 20 periods, were produced, to investigate antiferromagnetic coupling causing spin-flip events.

**Experimental details**

Ion-assisted magnetron sputter deposition was performed in a high-vacuum environment (approximately $5.6 \times 10^{-5}$ Pa or $4.2 \times 10^{-7}$ Torr) to deposit the Fe/Si and Fe/Si + $^{11}B_4C$ stacked multilayer samples (as illustrated in Figure 1) and the Fe/Si and Fe/Si + $^{11}B_4C$ multilayers with $\Lambda$ = 25 Å, $\Gamma$ = 0. And N = 20. The films were grown on single-crystalline Si (001) substrates measuring $10 \times 10 \times 1$ mm$^3$ with a native oxide layer. Sputtering targets included Fe (99.95% pure, 75 mm diameter), $^{11}B_4C$ (99.8% chemical purity, over 90% isotopic purity, 50 mm diameter), and Si (99.95% pure, 75 mm diameter). The magnetron sputter sources operated continuously, with computer-controlled shutters for each target material allowing precise control of atomic fluxes and enabling accurate multilayer deposition. The deposition rates were approximately 0.4 Å/s for Fe and Si, while $^{11}B_4C$ deposited at 0.08 Å/s. The $^{11}B_4C$ rate was determined using X-ray reflectivity (XRR) measurements and fitting of co-sputtered Fe/Si + $^{11}B_4C$ multilayers to extract period thicknesses. Substrates were maintained at ambient temperature (293 K) without intentional heating during deposition and were rotated at 8 rpm to ensure uniform film thickness. The initial 3 Å of deposition was performed with the substrate held at floating potential, followed by a -30 V substrate bias. To enhance the Ar-ion flux, a magnetic field aligned with the substrate normal was generated using a coil, condensing the plasma towards the substrate. Further details of the deposition system can be found elsewhere.[17]

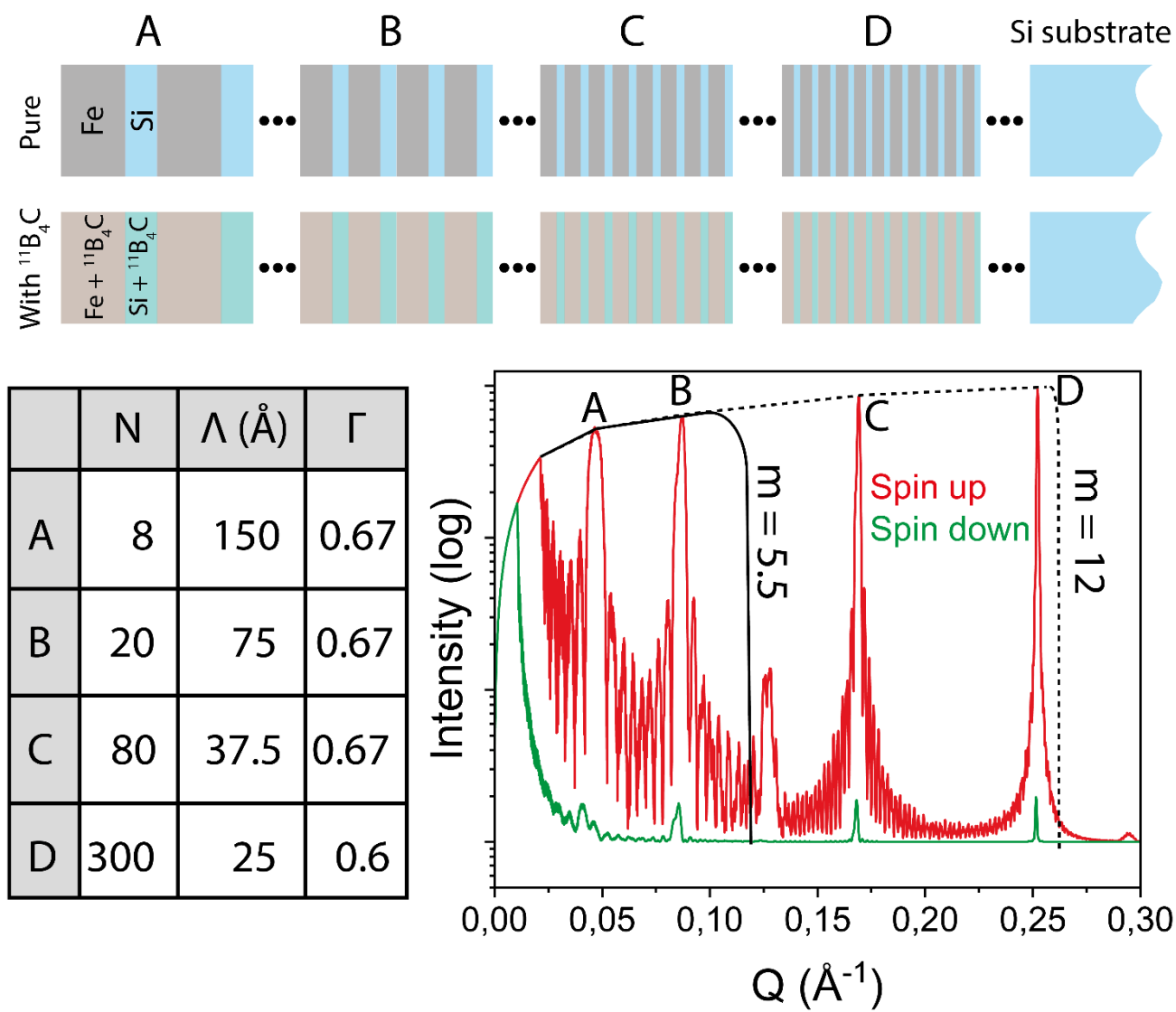

*Figure 1. Schematic drawing of the four sets of multilayers A+B+C+D for the two stacked samples with the number of periods( N), period thickness( Λ), and layer thickness ratio (Γ=iron layer thickness/Λ ) listed for each multilayer in the stack. The graph in the figure shows PNR simulations of Fe/Si + $^{11}B_4C$ stack ( the primary Bragg peaks for A, B, C and D are optimized for the highest possible intensity assuming 5 Å interface width). The solid black line represents state-of-the-art Fe/Si m = 5.5 supermirror[18], whereas the dashed line represents the possible reflectivity for a supermirror with m =12 using Fe/Si + $^{11}B_4C$.*

X-ray reflectivity measurements were performed on a Malvern Panalytical Empyrean diffractometer using Cu-Kα radiation and a PIXcel detector. The incident beam was equipped with a Göbel mirror and a 0.5° divergence slit, while the diffracted beam side utilized a parallel plate collimator with a 0.27° collimator slit. Multilayer periods were determined from the positions of the Bragg peaks in X-ray reflectivity measurements.

The wafer curvature method was employed to measure the internal stress in the multilayers by analyzing the bending of the substrate. X-ray diffraction (XRD) scans were performed and the change in curvature was determined using the Stoney equation to calculate the film stress. The diffractometer used was a Panalytical X'Pert diffractometer, using Bragg-Brentano geometry.

The 4 spin state neutron measurements have been performed on the SuperADAM[19] reflectometer at the Institute Laue-Langevin (ILL)[20]. This angle dispersive reflectometer provides a highly collimated fixed wavelength cold neutron beam of 5.2 Å with an energy spread of $\Delta\lambda = 0.5\%$. Incident polarization (99.87%) was achieved by a double-bounce bandpass Fe/Si mirror, while a standard m=3 Fe/Si supermirror analyzer with efficiency of 99.26% was used for polarization analysis. Each of the stacked multilayer samples, $20 \times 20$ mm$^2$ in size, were illuminated by a $1 \times 30$ mm$^2$ neutron beam. The perpendicular momentum transfer ($Q_z$) was adjusted by varying the angle of incidence of the neutron beam on the sample surface, and the reflected intensity was collected by a position-sensitive $^3$He detector. In polarized neutron reflectometry (PNR), neutron spin states, typically labelled as "up" or "down" relative to an applied magnetic field, enable the study of spin-dependent scattering. A four-state PNR measurement includes non-spin-flip (NSF) channels ("up-up" or "down-down"), where the spin remains unchanged, and spin-flip (SF) channels ("up-down" or "down-up"), where the spin reverses. These channels provide detailed information of the in-plane and out-of-plane magnetization profiles, enabling the separation of magnetic and nuclear scattering contributions for precise characterization of magnetic structures.

Two-spin state PNR measurements on Fe/Si and Fe/Si + $^{11}$B$_4$C, with $\Lambda = 25$ Å, $\Gamma = 0$. And N = 20, were conducted using the MORPHEUS instrument at the Swiss Spallation Neutron Source (SINQ) at the Paul Scherrer Institute (PSI). A polarized neutron beam with a wavelength of 4.825 Å was directed at the sample at small incidence angles ($\theta$), reflecting at each interface before detection by a $^3$He detector. Measurements were performed at external magnetic fields of approximately 0 mT (remanent field of 0.2 mT) and 40 mT, covering Q ranges of 0-0.03, 0.09-0.14, and 0.21-0.25 Å$^{-1}$. PNR is sensitive to the sample's spin-dependent scattering length density (SLD) and reveals information about the magnetization profile. The two spin states yield distinct curves in the PNR measurements, with Bragg peaks arising from constructive interference.

Magnetic properties were characterized at room temperature using vibrating sample magnetometry (VSM) in longitudinal geometry. VSM measurements were conducted over a field range of -1 T to 1 T to determine the samples' saturation magnetization and saturation field. The magnetic property measurements were performed using VSM (Vibrating Sample Magnetometer) option of Physical

Property Measurement System (PPMS: Quantum Design Japan, Inc.) at the CROSS user laboratory II in Japan.

X-ray and neutron reflectivity data were fitted to determine the multilayer periods, individual layer thicknesses, and interface roughness. Reflectivity fitting was performed using GenX,[21] which utilizes the Parratt recursion formalism to simulate and fit reflectivity data by accounting for each interface and calculating the total reflected intensity. According to Figure 1, each sample was modeled as a stack of four multilayers (A, B, C, and D), each multilayer consisting of multiple periods of uniform thicknesses.

For stacked multilayer samples, containing four sets of multilayers with a different period thickness and number of periods, as seen in Figure 1, each set generates a primary Bragg peak at a scattering vector governed by the modified form of Bragg's law.[22] Thicker periods result in Bragg peaks at lower vectors, while thinner periods give rise to peaks at higher vectors. Higher order Bragg peaks appear at multiples of the primary peak vector. Achieving high reflectivity, when dealing with thinner period thicknesses, requires a larger number of periods.

The m-value in neutron reflectometry indicates a multiple of the critical scattering vector, $Q_c$, below which there is a total external reflection of neutrons for a material.[23] For m = 1, this corresponds to the $Q_c$ where total external reflection occurs, while higher m-values (e.g., m = 2 , m = 3) represent multiples of this vector, often achieved through multilayer coatings that extend the reflection range. Higher m-values allow materials to reflect neutrons at larger vectors (higher Q-values), which is especially useful in applications like neutron supermirrors. Polarizing neutron optics are typical Fe/Si supermirrors, where the m value corresponds to the $Q_c$ of Ni and is also what is used for this study.

**Results and discussion**

First, we aim to demonstrate the amorphization of the multilayer, specifically the Fe layers, upon incorporating $^{11}B_4C$. This is achieved through X-ray diffraction (XRD) analysis. Figure 2(a) presents the XRD data, focusing on the range where the Fe (110) peak is expected. As observed, the Fe peak is absent in the $^{11}B_4C$-incorporated sample, confirming that $^{11}B_4C$ successfully amorphized the stacked multilayer. No other peaks other than the Si substrate peak is seen outside this range.

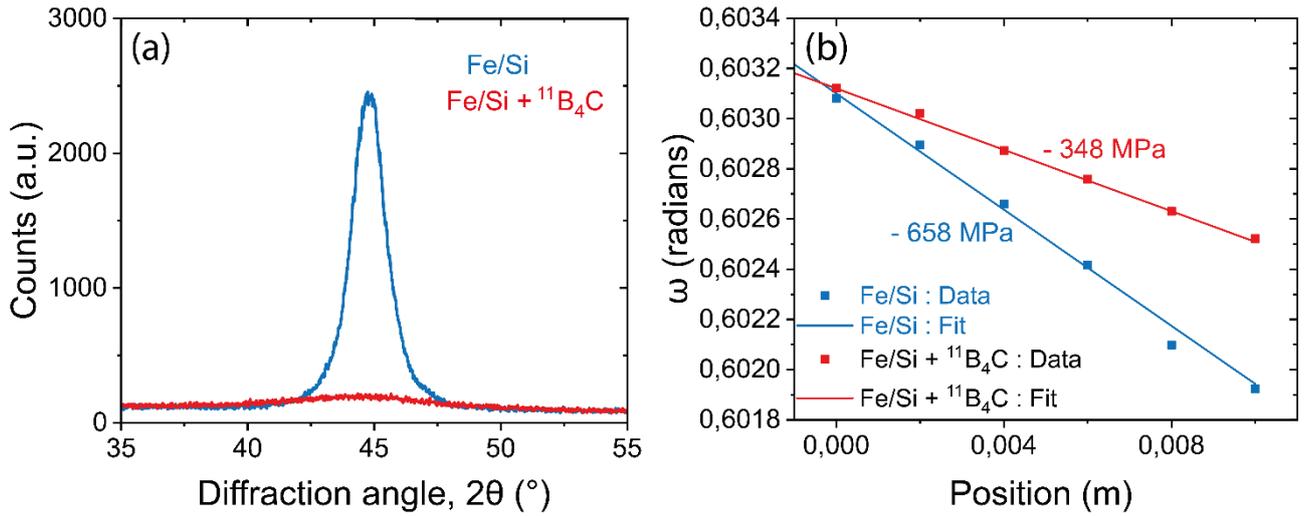

*Figure 2. X-ray diffraction. (a) X-ray diffraction focusing on the Fe (110) peak, for Fe/Si and Fe/Si $^{11}B_4C$ stacked multilayers. (b) Wafer-curvature measurements of the Fe/Si and Fe/Si + $^{11}B_4C$ stacked multilayers measured from the centre (0 mm) to the edge (8 mm) of the samples. A linear fit can be seen used to calculate the compressive/tensile stress using Stoney's equation.*

From wafer-curvature measurements, in Figure 2(b), of the two stacked multilayer samples, Fe/Si and Fe/Si + $^{11}B_4C$, it is evident that the incorporation of $^{11}B_4C$ significantly reduced stress from -658 MPa to -348 MPa, nearly halving the value. This stress reduction can be attributed to the amorphization effect, as amorphous films typically exhibit significantly lower stress compared to their crystalline counterparts. Such a decrease in stress enables the possibility of coating thicker multilayers, effectively addressing stress-related limitations that have previously restricted the achievement of m-values beyond m = 5.5.

The experimental verification of the four Bragg peaks shown in the simulation in Figure 1 would determine if a supermirror geometry is feasible up to m = 12, however not taking into account stress related limitations. The current state-of-the-art Fe/Si supermirrors achieve high reflectivity up to a scattering vector of Q = 0.12 Å$^{-1}$, corresponding to an m-value of 5.5.[18] This simulation takes into account the over-illumination of the sample but does not incorporate lateral impurities or accumulated roughness.

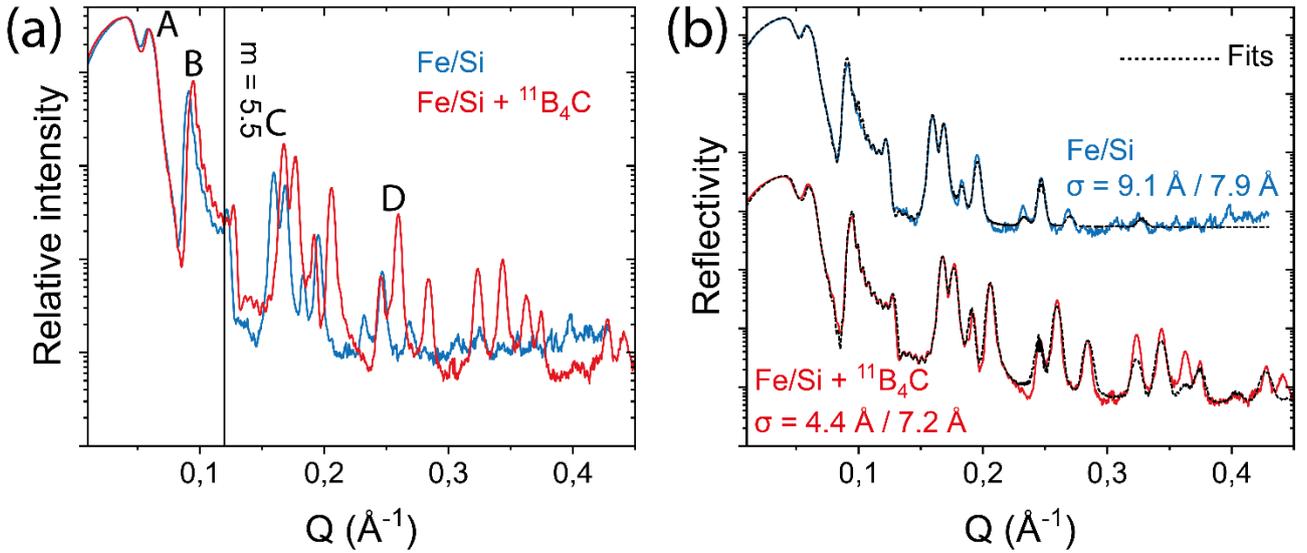

*Figure 3. X-ray reflectivity (XRR) data of Fe/Si and Fe/Si + $^{11}B_4C$ stacked multilayers, where the intensity is in log-scale. (a) shows the normalized data while (b) shows the vertically shifted data along with fits. The solid line in (a) indicates the state-of-the-art m = 5.5 limit. The interface width, σ, is described as X Å / Y Å, where X represents the interface on top of the Fe layer and Y the interface width on top of the Si layer. The interface width values are averaged from all the sets in the stack.*

In Figure 3, we evaluated the interface quality of the Fe/Si and Fe/Si + $^{11}B_4C$ stacked multilayer samples through XRR and analyzed the corresponding fits. In Figure 3(a), the normalized data of the two samples are shown for a qualitative comparison. In panel (b), the data are vertically shifted to highlight the scattering vector (Q) where the Bragg peaks fade into the background, making it easier to distinguish the fits of the two samples. What is apparent in (a) is that all Bragg peaks are significantly higher in intensity for Fe/Si + $^{11}B_4C$ compared to Fe/Si. Furthermore, additional Bragg peaks beyond 0.3 Å$^{-1}$ are visible for Fe/Si + $^{11}B_4C$, which are absent for Fe/Si, clearly indicating the improved interface quality for Fe/Si + $^{11}B_4C$ compared to the pure Fe/Si. The fits show that the average interface width on top of the Fe layers was reduced to less than half compared to Fe/Si, while the interface width above the Si layers also decreased, although less dramatically. This reduction in interface width enhances the XRR. The significant differences between the Bragg peak intensities in the PNR simulation in Figure 1 and the XRR data in Figure 3 is primarily due to the differences between X-ray and neutron scattering. The interface quality and consequently the neutron reflectivity performance can first be assessed through X-ray reflectivity.[24–27] However, X-ray and neutron reflectivity differ in their optical responses, as X-rays interact with the electron cloud, whereas neutrons interact with atomic nuclei. This difference affect the interpretation of each layer's properties. Both techniques rely on the scattering length density (SLD) to determine the theoretical reflective potential, though the SLD varies based on whether X-rays or neutrons are used, depending on the mass density and interaction characteristics of each probe.[28]

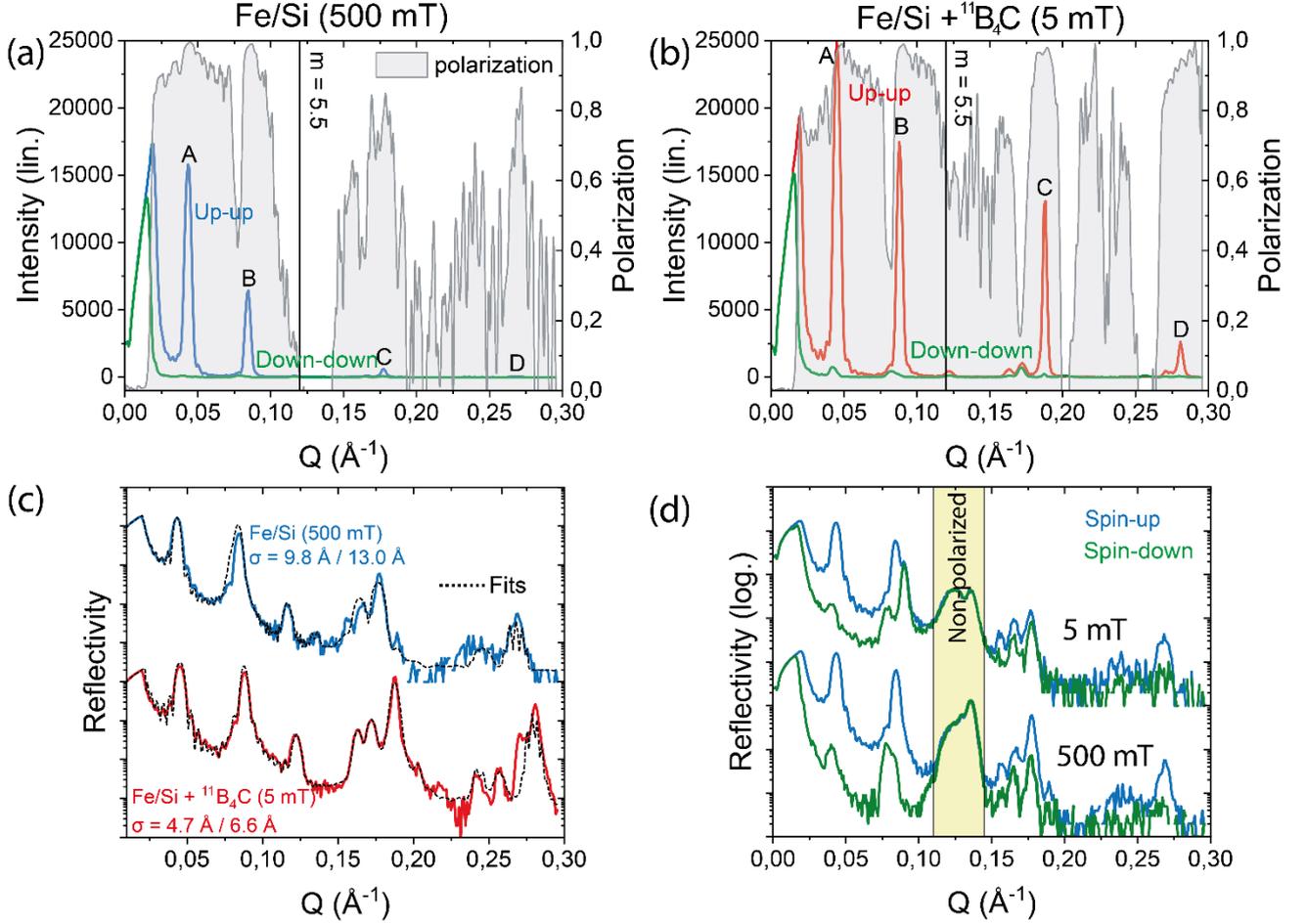

*Figure 4. Polarized neutron reflectivity (PNR) with polarization analysis of (a) Fe/Si and (b) Fe/Si + $^{11}B_4C$ stacked multilayers. The grey areas indicate the degree of polarization. The solid line indicates the current state-of-the-art m = 5.5 limit. The vertically shifted up-up reflectivity fits are shown in (c). Note that in (a-c) shows Fe/Si measured at 500 mT and Fe/Si + $^{11}B_4C$ measured at 5 mT. (d) shows the vertically shifted spin-up (up-up + up-down) and spin-down (down-down + down-up) reflectivities of the Fe/Si stacked multilayer sample at 5 mT and 500 mT. The yellow region marks the non-polarized region.*

In Figure 4, neutron reflectivity measurements for the two stacked multilayer samples, Fe/Si and Fe/Si + $^{11}B_4C$, are shown for up-up and down-down reflectivity curves. Measurements were taken at 5 mT and 500 mT, assuming that 500 mT would saturate both samples while 5 mT would only saturate the $^{11}B_4C$-incorporated sample. In panel (a), the Bragg peaks at positions C and D are barely visible due to large interface widths, resulting in low reflectivity. This is consistent with that state-of-the-art polarizing supermirrors typically do not achieve reflectivity beyond Q = 0.12 Å$^{-1}$. In panel (b), the reflectivity for the primary Bragg peaks significantly increases with the addition of $^{11}B_4C$: A by 54%,

B by 172%, C by 2170%, and D by 4679%, demonstrating a remarkable improvement in reflectivity. This increase in reflectivity is attributed to the decrease in interface width when $^{11}B_4C$ is incorporated. Additionally, the Bragg peaks at C and D are clearly visible in (b), making it feasible to achieve reflectivities with a supermirror geometry up to a scattering vector Q of 0.26 Å$^{-1}$, corresponding to m = 12. A comparison of the interface widths obtained from XRR and PNR reveals a consistent decrease in interface width for both Fe-on-Si and Si-on-Fe interfaces upon incorporating $^{11}B_4C$.

Further, the polarization of the $^{11}B_4C$ incorporated sample is polarized in broader regions than the Fe/Si sample, which confirms the superior polarization, as shown in the grey areas in Figure 4(a-b), indicating a more pronounced and consistent polarization.

The spin-up and down (including the spin-flip) of the PNR measurements for the Fe/Si sample can be seen in Figure 4(d), where there is a broad non-polarized peak/s between 0.11 and 0.145 Å$^{-1}$ for the Fe/Si sample, regardless of whether the applied field is at 5 mT or 500 mT. Since the position of this peak is at half the scattering vector value of the multilayer with the thinnest multilayer period (multilayer D) this suggests that this region could have an antiferromagnetic half-order Bragg peak caused by the multilayer set with the thinnest periods, D.

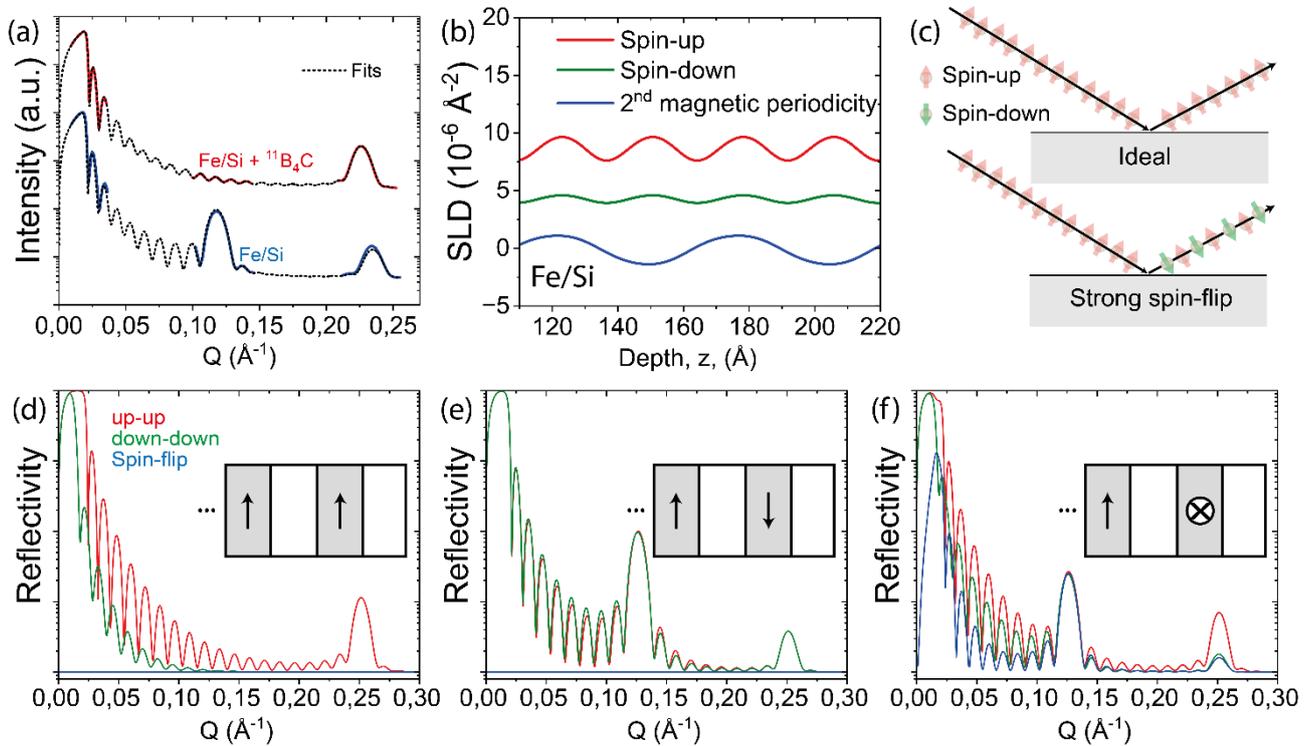

*Figure 5. Polarized neutron reflectivity measurements and corresponding fits of the Fe/Si sample with Λ = 25 Å, Γ = 0.5, and N = 20, measured at 40 mT shown in (a). (b) shows the SLD profile of the Fe/Si multilayer stemming from the 40 mT fit. (c) shows a schematic of the ideal reflection and a reflection which suffers from*

*spin-flip events. (d-f) Four-spin-state PNR simulations of the Fe/Si sample in (a) with every second magnetic layer having different magnetization direction; (d) 0°, (e) 180° and (f) 90°.*

Polarized neutron reflectivity measurements on the periodic Fe/Si multilayers with periodicity $\Lambda = 25$ Å, as shown in Figure 5(a), reveal the presence of an antiferromagnetic peak, appearing at half the scattering vector of the ordinary primary Bragg peak.[29,30] The antiferromagnetic half-order Bragg peak is present even though the external field is 40 mT, indicating strong antiferromagnetic coupling resistant to high external fields. For the stacked Fe/Si multilayer, this half-order Bragg peak will cause a high spin-flip intensity at $Q = 0.13$ Å$^{-1}$, matching the periodicity of the multilayer set with a periodicity $\Lambda = 25$ Å.

The scattering length density profiles for the 40 mT measurement in Figure 5 (b), obtained from reflectivity fitting, indicates a magnetic periodicity of approximately 50 Å, twice the thickness of the spin-up and spin-down periodicity (25 Å). The positive peaks resembles a net magnetization in that layer which has a vector component directed towards the external field while the negative valleys represents a net magnetization with a vector component directed against the external field. When comparing the PNR results for Fe/Si with the Fe/Si multilayers incorporating $^{11}B_4C$, as shown in Figure 5 (a), a clear difference emerges. The Fe/Si + $^{11}B_4C$ multilayer shows no antiferromagnetic half-order Bragg peak and exhibits clear polarization at the first-order Bragg peak at 40 mT. This demonstrates that the inclusion of 15 vol.% $^{11}B_4C$ eliminates antiferromagnetic coupling, creating a magnetic periodicity equal to twice the periodicity of the bilayer thickness. Thus, $^{11}B_4C$ effectively suppress antiferromagnetic coupling between magnetic layers.

Figure 5(d-f) shows that the polarization gets worse as soon as the angle of every second magnetic layer is canting. The spin-flip appear as soon as the angle between magnetic moments in adjacent layers is neither parallel with nor against the external field. The 180° (b) may not have spin-flip but there is a half-order Bragg peak and neither the half-order nor 1$^{st}$ order Bragg peak are polarized, meaning that half of the neutrons have the undesired spin state.

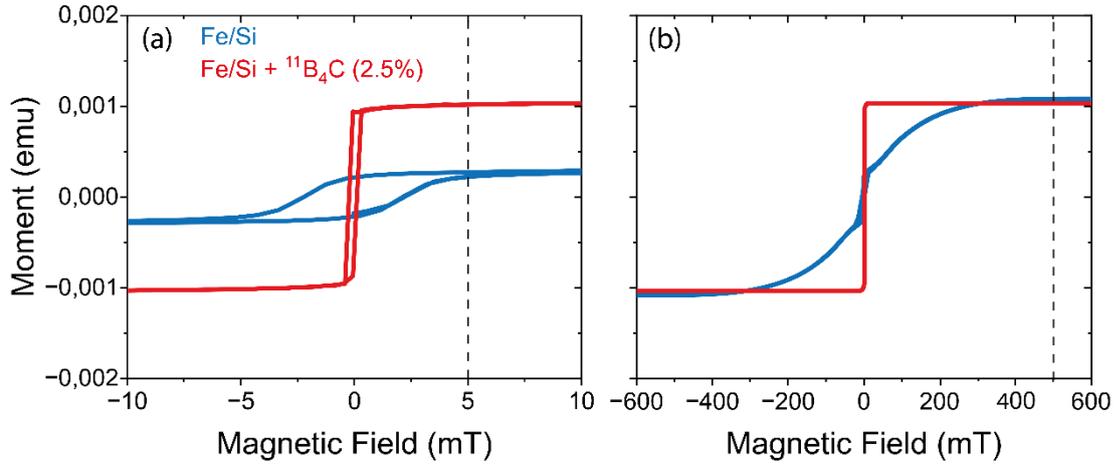

*Figure 6. Vibrating sample magnetometry (VSM) of Fe/Si and Fe/Si + $^{11}B_4C$ multilayer sample with N = 20, $\Lambda$ = 25 Å and $\Gamma$ = 0.5 showing the range from (a) -10 mT to 10 mT and (b) -600 mT to 600 mT in. The dashed lines indicate the 5 and 500 mT magnetic fields which were the fields used to measure the 4-spin-state PNR.*

Vibrating sample magnetometry (VSM) data in Figure 6 (a-b) further shows the magnetic behaviour of the Fe/Si and Fe/Si + $^{11}B_4C$ multilayers, both with $\Lambda$ = 25 Å, $\Gamma$ = 0.5, and N = 20. For Fe/Si, the magnetization does not saturate even at 5 mT, as indicated by the slope-like behaviour in the hysteresis loop, a common signature of antiferromagnetic coupling where every second magnetic layer is ordered differently than every other magnetic layer.[31] In such magnetically coupled multilayers, the interlayer exchange coupling may be strong enough to resist saturation until external fields of several hundreds of mT, as observed here. As seen in Figure 6 (b), Fe/Si saturates at 600 mT, while Fe/Si + $^{11}B_4C$ saturates at just 1 mT, as observed in Figure 6 (a). This behaviour explains why high-intensity antiferromagnetic half-order Bragg peaks are observed in both the 5 mT and 500 mT two-spin-state measurements (Figure 4(d)), and also understood in from the spin-flip intensities in the four-spin-state measurements shown in Figure 7.

To confirm whether the reflected neutrons are indeed spin-up neutrons as intended or if any spin-flip phenomena occur, four-spin-state PNR measurements are required. Investigating spin-flip phenomena using four-spin-state PNR requires both a polarizer, to prepare the neutron spin-state before interaction, and an analyzer, to determine the spin-state after interaction. Spin flip events can occur due to magnetic domains,[12] or antiferromagnetic/biquadratic coupling between magnetic layers, causing the magnetic moments in every or every second magnetic layer to be misaligned. This phenomenon is undesired in polarizing neutron optics since it leads to a worse polarization of the beam, making it important to minimize or eliminate spin-flip behaviour.

In both ferromagnetic and antiferromagnetic systems, the spin-flip scattering intensity is sensitive to the components of the magnetic moments perpendicular to the scattering vector Q. In the specular reflectivity measurements performed here, the scattering vector Q is oriented perpendicular to the layers, and thus the local magnetic moments causing the spin-flip events are oriented in the plane of the layers. If the average magnetization of a layer is canted at an angle relative to the external magnetic field, rather than aligned parallel to it, spin-flip phenomena will occur, thereby diluting the neutron beam polarization. In Figure 4(d), it is evident that Fe/Si multilayers exhibit high-intensity spin-down peaks at the Bragg positions at 5 mT, which can be attributed to magnetic domains. At a saturation field of 500 mT, these domains are aligned with the magnetic field, however, interlayer coupling between layers still persists, causing spin-flip events.

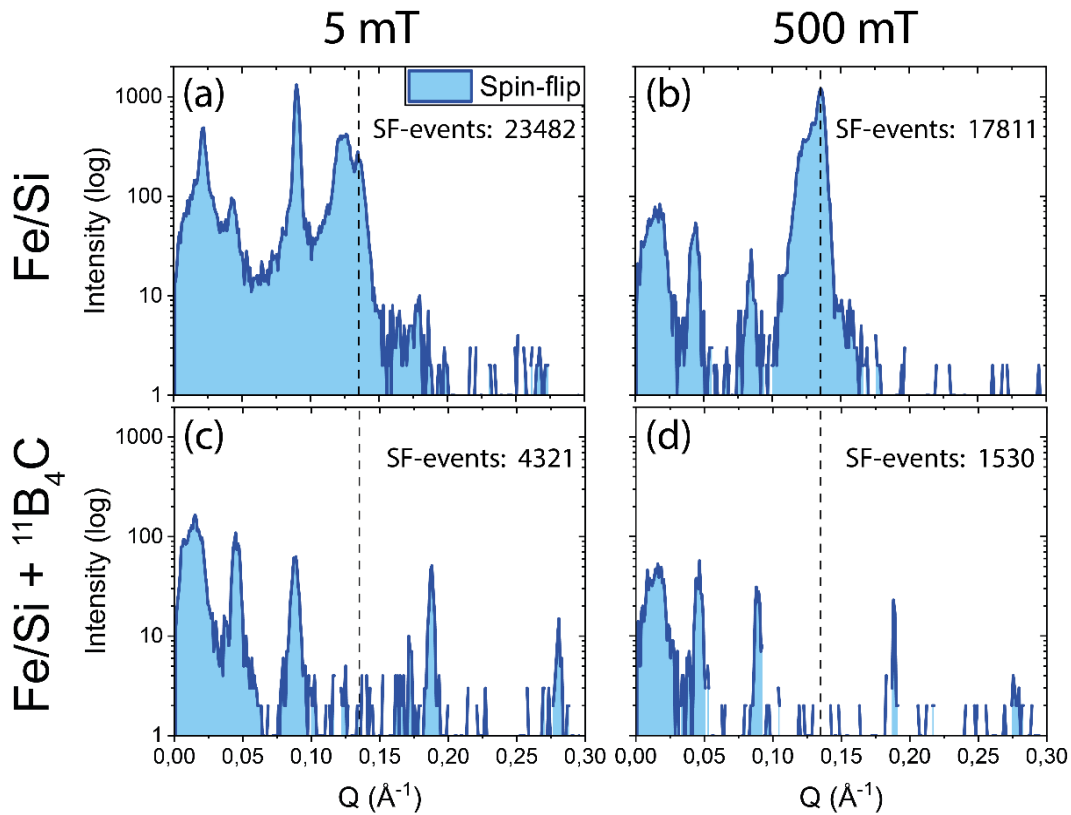

*Figure 7. Spin-flip intensities (up-down and down-up) for the two stacked multilayer samples: (a) Fe/Si at 5 mT, (b) Fe/Si at 500 mT, (c) Fe/Si + $^{11}B_4C$ at 5 mT, and (d) Fe/Si + $^{11}B_4C$ at 500 mT. The total number of spin-flip events is listed for each experiment.*

Figure 7 shows the spin-flip intensities (up-down + down-up) for Fe/Si and Fe/Si + $^{11}B_4C$ stacked multilayers at two different fields 5 mT and 500 mT. These intensities arise only when the intended neutron spin has flipped due to interaction with magnetic inconsistencies, such as misaligned domains or coupling that deviate from the direction of the externally applied magnetic field. Spin-flip intensities are undesirable since they reduce the polarization purity of the neutron beam. Figure 7 (a-

b) shows the spin-flip intensities for the Fe/Si sample at 5 mT, where the sample is not fully magnetically saturated, and 500 mT, where saturation is expected. Figure 7 (c-d) show corresponding measurements for the Fe/Si + $^{11}B_4C$ sample. Comparing the spin-flip intensities between the two samples, the Fe/Si + $^{11}B_4C$ sample consistently show significantly lower spin-flip intensities, particularly at the Bragg positions and the region between 0.1 Å$^{-1}$ and 0.15 Å$^{-1}$. Notably, the Fe/Si + $^{11}B_4C$ sample at 5 mT experience substantially less spin-flip intensity than Fe/Si at 500 mT, clearly showing the substantial reduction of spin-flip scattering when incorporating $^{11}B_4C$. This reduction is expected since the addition of $^{11}B_4C$ amorphizes the Fe layers, making the Fe magnetically soft,[15] and the inherent magnetic domains causing spin-flip scattering are eliminated.

As noted, in the case of spin-flip, a half-order Bragg peak at Q = 0.13 Å$^{-1}$, indicative of antiferromagnetic or biquadratic coupling, remains present at higher external fields. This observation is in good agreement with the VSM data, shown in Figure 6, as well as the PNR results, shown in Figures 4(d) and 5(a), which confirm that the sample is not saturated, and the half-order Bragg peak remains regardless of the applied magnetic field. A visualization of the probable biquadratic coupling in Figure 5(d-f) shows that when the sample's magnetization has a component perpendicular to the neutron polarization axis, spin-flip scattering occurs. Differences between these intensities may indicate asymmetric magnetic structures or inhomogeneities, such as unequal domain sizes or variations in magnetic coupling across layers. The spin-flip events were decreased with 82% when 5 mT was applied while decreasing 91% when 500 mT was applied. The remaining spin-flip intensities observed in the measurements, Figure 7 (c-d), may be attributed to "spin leakage" from the instrument itself. This effect likely arises from imperfections in the polarizer and analyzer optics used during the experiment. Such instrumental limitations can lead to incomplete polarization or analysis of the neutron beam and are not indicative of intrinsic magnetic properties or spin-dependent scattering from the investigated Fe/Si + $^{11}B_4C$ multilayer. An instrumental spin leakage is supposed to be solved using $^{11}B_4C$ incorporation. Thus, the spin-flip can be argued to be decreased significantly beyond what is reported when incorporating $^{11}B_4C$.

**Conclusion**

The incorporation of $^{11}B_4C$ into Fe/Si multilayers significantly enhances the performance of polarizing neutron optics. X-ray reflectivity and polarized neutron reflectivity measurements revealed significantly improved interface quality, evidenced by the appearance of additional Bragg peaks and substantially more intense Bragg peaks compared to pure Fe/Si multilayers. Furthermore, the inclusion of $^{11}B_4C$ improved polarization performance, and drastically reduced spin-flip scattering, even at a low external field of 5 mT. Notable is that Fe/Si + $^{11}B_4C$ at 5 mT outperformed Fe/Si at 500 mT. This improvement is attributed to the amorphization effect of $^{11}B_4C$, which disrupts crystallinity of Fe layers. This eliminates magnetic anisotropy and strong interlayer interactions such as antiferromagnetic or biquadratic coupling, making the magnetic moments in Fe layers easier to saturate and control. Unlike pure Fe/Si, which exhibited strong spin-flip scattering at Q = 0.13 Å$^{-1}$ due to interlayer coupling effects, Fe/Si + $^{11}B_4C$ showed no such phenomena. Additionally, the incorporation of $^{11}B_4C$ halved the internal stress of the stacked multilayer, enabling the possibility of fabrication of thicker films by addressing stress-related limitations for higher m-values which requires additional periods. These enhancements demonstrate that Fe/Si + $^{11}B_4C$ multilayers significantly outperform pure Fe/Si in reflectivity, polarization efficiency, suppression of spin-flip scattering, and stress, enabling the development of high-performance polarizing supermirrors with higher m-values.

**Acknowledgements**

This work was supported by the Swedish Research Council (VR), grant number 2019-04837_VR. A.Z acknowledge the grant 2022-D-03 from the Hans Werthén Foundation, the grant from Royal Academy of Sciences Physics grant, PH2022-0029, the Lars Hiertas Minne foundation grant FO2022-0273 and the Längmannska Kulturfonden grant BA23-1664. We are grateful for the SuperADAM beamtime (CRG-3036) at Institut Laue-Langevin (ILL). For the magnetic property measurements we are thankful to the CROSS user laboratory II in Japan. We are also thankful to Dr. Motoyuki Ishikado for his kind support.